\documentclass[runningheads]{llncs}

\usepackage[T1]{fontenc}
\usepackage{graphicx}
\usepackage{subcaption}
\usepackage{url}
\usepackage{multirow}
\usepackage{booktabs}
\usepackage{tabularx} % 表の幅を調整するために必要
\usepackage{amsmath}
\usepackage{mathtools}
\usepackage{vtable,hhline}
\usepackage{amsmath, amssymb}
\usepackage{enumitem}
\usepackage{ascmac}
\usepackage{array}
\usepackage{longtable}
\begin{document}
\title{Relationship Analysis of Image-Text Pair \\ in SNS Posts}
\titlerunning{Relationship Analysis of Image-Text Pair}
% If the paper title is too long for the running head, you can set
% an abbreviated paper title here
%
% \author{Takuto Nabeoka\inst{1}\orcidID{} \and
% Yijun Duan\inst{2}\orcidID{} \and
% Qiang Ma\inst{2}\orcidID{}}
\author{Takuto Nabeoka\inst{1} \and
Yijun Duan\inst{2} \and
Qiang Ma\inst{2}}
\authorrunning{T. Nabeoka et al.}
% First names are abbreviated in the running head.
% If there are more than two authors, 'et al.' is used.
%
\institute{Graduate School of Informatics, Kyoto University\\
% Yoshida-Honmachi, Sakyo-ku, Kyoto-shi, Kyoto 606--8501, Japan \and
\email{nabeoka.takuto.x79@kyoto-u.jp}
\and 
Graduate School of Science and Technology, Kyoto Institute of Technology\\
% Matsugasaki Hashigami-cho, Sakyo-ku, Kyoto-shi, Kyoto 606--8585, Japan
\email{\{yijun,qiang\}@kit.ac.jp}}
\maketitle              % typeset the header of the contribution
\begin{abstract}
% The abstract should briefly summarize the contents of the paper in
% 150--250 words.
% Social networking services (SNS) contain a large number of image-text posts, 
% making it crucial to analyze the relationship between images and text for effective information retrieval and understanding. 
% In this study, 
% we focus on the classification of image-text pairs in SNS. 
% Previous studies have struggled to distinguish relationships beyond similarity, 
% highlighting the need for improved methods. 
% To address this issue, 
% we propose a graph-based model that considers both intra- and inter-pair relationships to classify similarity and complementarity. 
% Specifically, 
% we embed images and text using CLIP, 
% perform separate clustering, 
% and construct an ITRC-Line Graph by treating each cluster as a node and swapping edges and nodes in a pseudo-graph representation. 
% We then employ a Graph Convolutional Network (GCN) to learn node and edge representations. 
% Finally, 
% we fuse the learned representations with the original image-text embeddings to classify their relationships. 
% Experimental evaluations using a publicly available dataset demonstrate the effectiveness of our approach.
Social networking services (SNS) contain vast amounts of image-text posts, 
necessitating effective analysis of their relationships for improved information retrieval. 
This study addresses the classification of image-text pairs in SNS, 
overcoming prior limitations in distinguishing relationships beyond similarity. 
We propose a graph-based method to classify image-text pairs into similar and complementary relationships.
Our approach first embeds images and text using CLIP, 
followed by clustering.
Next, 
we construct an Image-Text Relationship Clustering Line Graph (ITRC-Line Graph), 
where clusters serve as nodes.
Finally,
edges and nodes are swapped in a pseudo-graph representation. 
% Our approach embeds images and text using CLIP, 
% performs clustering,
% and constructs an Image-Text Relationship Clustering Line Graph (ITRC-Line Graph), 
% where clusters serve as nodes, 
% and edges and nodes are swapped in a pseudo-graph representation. 
A Graph Convolutional Network (GCN) then learns node and edge representations, 
which are fused with the original embeddings for final classification. 
Experimental results on a publicly available dataset demonstrate the effectiveness of our method.
\keywords{SNS \and Multimodal Learning \and Clustering \and GNN \and Relationship Analysis.}
\end{abstract}
\section{Introduction}
\label{sec:intro}
The widespread use of social networking services (SNS), 
such as X (formerly Twitter), 
has made it easier for anyone to send and receive information.
A significant proportion of SNS posts contain images, 
which are visually engaging and tend to attract greater attention. 
Studies indicate that approximately 42\% of posts on X include images, 
receiving 22.8\% more engagement than text-only posts \cite{lee2015What}. 
This highlights the importance of both images and text in SNS and underscores the necessity of analyzing their relationships to better understand information dissemination and optimize SNS utilization.
% 近年，
% X(旧Twitter)等のソーシャル・ネットワーキング・サービス(以下，SNS)の普及により，
% 誰でも手軽に情報の送受信が可能である．
% SNSでは画像付き投稿が多く，
% 画像は視覚的に目を惹き，
% 関心を集める傾向がある．
% 実際にXにおける投稿の約42\%が画像を含み，
% 画像なしの投稿に比べて約 22.8\%多くエンゲージメントを獲得しているという報告もある\cite{lee2015What}．
% このことから，
% 画像とテキストはSNSを構成する重要な要素であり，
% それらの関連性を分析することは情報の拡散やSNSの有効活用の理解に寄与する．

Analyzing image-text relationships enables the efficient extraction of relevant information from large datasets, 
facilitating various applications. 
For instance, 
it can support content recommendation in post-assistance and document creation systems, 
contribute to the development of multimodal datasets, 
and enhance downstream tasks such as remote sensing \cite{yuan2022Exploring}.
% and tourism promotion \cite{tu2023Image}. 
% Particularly in SNS, 
% such analyses are crucial for the rapid classification of large-scale information, 
% aiding in the detection of inappropriate content and spam, 
% as well as improving the efficiency of information collection during disaster events.
% 画像ーテキスト対の関連性分析は，
% 膨大な情報の中から目的に応じた情報を抽出することで，
% 多様なタスクに応用可能である．
% 例えば，
% 投稿支援や資料作成システムにおける適切なコンテンツ推薦，
% マルチモーダルデータセットの構築，
% 更にはリモートセンシング \cite{yuan2022Exploring} や観光プロモーション \cite{tu2023Image} 等の下流タスクに活用されると考えられる．
% 特にSNS投稿における関連性分析は，
% 大量の情報を迅速に分類する必要がある場面で特に有用であり，
% 不適切コンテンツやスパムの検出，
% 災害時の情報収集の効率化などに貢献する．

\begin{figure}[t]
  \centering
  \includegraphics[width=\linewidth]{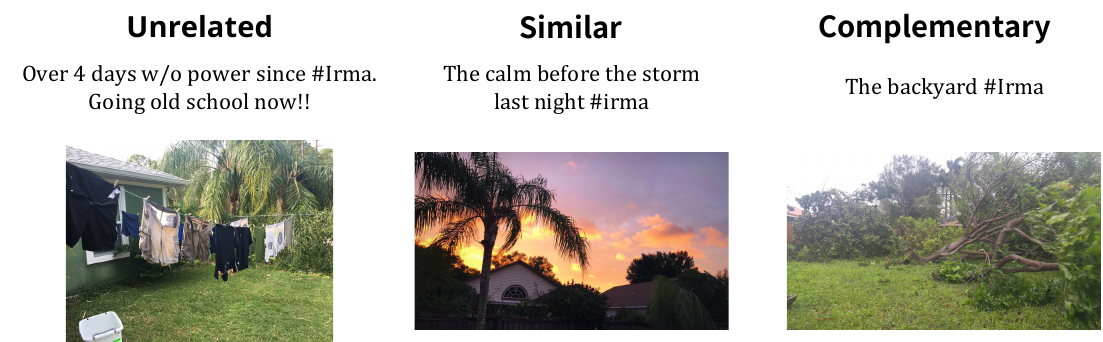}
  \caption{Example of X's posts and the image-text pair class defined in DisRel \cite{sosea2021Using}}
  % \caption{DisRel\cite{sosea2021Using}におけるXの投稿と関連性の分類の例}
  \label{fig:fig_relatedexample}
\end{figure}

% As a study analyzing the relationship between images and text, 
Sosea et al.\cite{sosea2021Using} developed the DisRel dataset from 4,991 posts collected from X during U.S. disasters in 2017. 
The posts are manually classified into three categories: ``Unrelated'', ``Similar'', and ``Complementary'' (Figure \ref{fig:fig_relatedexample}).
In the ``Unrelated'' example, 
an image of laundry and a text about going to school are unrelated. 
The ``Similar'' example shows an image representing ``the calm before the storm,'' 
sharing a clear relationship with the text. 
The ``Complementary'' example pairs a text about a backyard with an image of a backyard, 
providing additional context.

Among the different types of relationships, 
the ``Complementary'' relationship plays a crucial role in accurately understanding information, 
as it enhances comprehension through the integration of images and text.
% For example, 
% in the ``Complementary'' relationship case shown in Figure \ref{fig:fig_relatedexample}, 
% the text conveys that a tree in the backyard has fallen due to Hurricane Irma\footnote{A Category 5 hurricane that caused extensive damage across the United States and Caribbean nations in September 2017. Reference: \url{https://www.bbc.com/news/world-latin-america-41182991} (Accessed: \today)}, 
% and this information is reinforced when combined with the corresponding image.
However, 
``Complementary'' relationships in image-text pairs are rarely explicitly stated within each modality, 
making them difficult to detect automatically. 
In fact, the privious study \cite{sosea2021Using} reports that 
while the best-performing model achieved an F1-score of 0.82 for identifying ``Similar'' relationships, 
it only reached 0.62 for ``Complementary'' relationships.
Therefore, 
improving the classification accuracy of ``Complementary'' relationships remains a challenge. 
% Since ``Similar'' relationships share common information, 
% classification can be effectively achieved using image recognition techniques, 
% such as image captioning, 
% or multimodal models. 
% In contrast, 
% ``Complementary'' relationships require both image and text to fully convey the intended information, 
% necessitating a broader approach to capturing their relationships.
Since ``Similar'' relationships share common information, 
they can be effectively classified using image recognition or multimodal models. 
In contrast, 
``Complementary'' relationships require both image and text to capture their full meaning.
% 関連性の中で``補足(Complementary)'' 関係は，
% 画像とテキストの統合によって情報理解が深まるため，
% 正確な情報把握に重要である．
% 例えば図\ref{fig:fig_relatedexample}の補足関係の投稿では，
% テキストの内容は画像と組み合わせることで，
% イルマ\footnote{2017年9月にアメリカやカリブ海諸国に甚大な被害をもたらした最大カテゴリー5の超大型ハリケーン．参考:\url{https://www.bbc.com/japanese/41183701}(最終閲覧日:\today)}
% により裏庭の木が倒れる被害が伝わる．
% しかし，
% 画像ーテキスト対の補足関係は各モダリティに直接的に明示される例が少なく，
% 自動判別が困難である．
% 実際先行研究\cite{sosea2021Using}においても，
% 最も性能が良いと報告されたモデルで類似関係の判別のF1値が0.82であるのに対し，
% 補足関係では0.62に留まっている．
% そのため補足関係の判別精度向上が課題となっている．
% 類似関係は共有情報があるため，
% 画像キャプション生成等の画像認識技術や，
% マルチモーダルモデルの利用により分類性能が担保できると考えられる．
% 一方，
% 補足関係は画像とテキストの両者が揃って初めて完全な情報が伝達可能であるため，
% 関係性をより広範に捉える必要があると考えられる．

This study aims to improve the accuracy of capturing ``Complementary'' relationships in image-text pair classification, 
in addition to ``Similar'' relationships. 
% Traditional methods primarily classify image-text pairs based on their internal information but do not consider relationships between pairs. 
% To address this limitation, 
We propose a graph-based model that incorporates both intra-pair and inter-pair relationships.
First, 
we construct a heterogeneous graph, 
where each image-text pair is represented as an edge. 
This edge representation captures internal relationships within pairs. 
However, 
this graph tends to be sparse, 
making it difficult to fully leverage inter-pair information.
To overcome this issue, 
we introduce the ITRC-Graph (Image-Text Relationship Clustering Graph), 
where clusters of images and texts are treated as nodes. 
This graph structure enables the consideration of relationships between pairs through its edge representation.
Furthermore, 
to effectively update edge representations, 
we employ the Graph Convolutional Network (GCN) \cite{kipf2017SemiSupervised}. 
% a type of Graph Neural Network (GNN). 
Since GCN is designed for node representation learning, 
we transform the ITRC-Graph into the ITRC-Line Graph (Image-Text Relationship Clustering Line Graph) 
by swapping nodes and edges. 
This allows the GCN model to learn meaningful edge representations.
Finally, 
we fuse the updated edge representations 
with the original embeddings of each image-text pair 
and train a classifier to predict relationships efficiently using multiple sources of information.
Our proposed method captures not only internal information within pairs 
but also inter-pair relationships, 
improving the classification of both similar and complementary relationships in image-text pairs.
% 本研究では，
% 画像ーテキスト対の関連性分類において，
% 類似関係に加えて補足関係の捕捉精度を向上させることを目的とする．
% 従来の手法では，
% 各画像ーテキスト対内部の情報に基づいた分類が主流であったが，
% ペア間の関係性を考慮できていない．
% そこで本研究では，
% ペア内部の情報に加え，
% 他のペアとの関係性を考慮するためのグラフベースのモデルを提案する．
% まず各ペアをエッジとするヘテログラフを構築し，
% エッジ表現を利用することでペア内部の関係を捉える．
% しかし，
% このグラフはスパースであり，
% ペア間の情報を十分に活用できないという課題がある．
% そこで，
% 各画像やテキストをクラスタリングした各クラスターを各ノードとみなす
% ITRC-Graph（Image-Text Relationship Clustering Graph）を構築する．
% このグラフのエッジ表現を用いることで，
% ペア間の関係も考慮可能となる．
% 更に，
% グラフのエッジ表現を効果的に更新するため，
% GNNの一種である畳み込みグラフニューラルネットワーク(以下，GCN)\cite{kipf2017SemiSupervised}ベースのモデルを適用する．
% しかし，
% GCNはノードの表現学習をするため，
% ITRC-Graphのエッジとノードを入れ替えたITRC-Line Graph(Image-Text Relationship Clustering Line Graph)に変換し，
% GCNモデルを用いて学習を行う．
% 最終的に得られたエッジ表現と，
% 各画像ーテキスト対の元のエンベディングを融合して分類器を学習することで，
% 複数の情報を効率良く利用して関連性を予測する．
% 本手法によりペア内部の情報だけでなく，
% ペア間の関係性を考慮して．
% 画像ーテキスト対の類似関係と補足関係の関連性を捕捉することを目指す．

The contributions of this study are as follows:

% \begin{enumerate}
% \item We propose a model that utilizes the ITRC-Line Graph, 
% which is constructed by clustering images and texts, 
% and introduce a method for learning representations using GCN (Section \ref{subsec:Clustered Edge Embedding}).
% \item We present a method for fusing information from images, text, 
% and the ITRC-Line Graph to classify image-text pair relationships (Section \ref{subsec:Fused_Embedding}).
% \item We evaluate the proposed model through experiments using a public dataset (Section \ref{sec:Experiment}) and analyze the effectiveness of the graph-based approach (Section \ref{sec:Analysis}).
% \end{enumerate}

\begin{enumerate}
    \item We propose a novel task of classifying image-text pairs into similar and complementary relationships using graph-based representations, 
    facilitating applications of graph-based methods.
    
    \item We propose a method to construct the ITRC-Line Graph and learn representations using a Graph Convolutional Network (GCN).
    We also develop a fusion method that combines image, text, and ITRC-Line Graph information to improve classification accuracy (Section \ref{sec:method}).

    \item We validate our model through experiments on a public dataset (Section \ref{sec:Experiment}) 
    and analyze the impact of the graph-based approach (Section \ref{sec:Analysis}).
    Experimental results demonstrate that the proposed model effectively classifies image-text relationships, 
    especially in the "Complementary" category.
\end{enumerate}

% 本研究の貢献は以下の通りである．
% \begin{enumerate}
%     % \item 画像ーテキスト対の内部と外部の関係を考慮して，画像ーテキスト対の関連性を捕捉するグラフベースの手法を提案する
%     % \vspace{-0.5em}
%     \item 各画像とテキストをクラスタリングを利用して構成するITRC-Line Graphを用いたモデルを提案し，
%     GCNを用いて表現学習する手法を示す(\ref{subsec:Clustered Edge Embedding}節)．
%     \item 分類器で関連性を分類するために，画像，テキスト，ITRC-Line Graphからの情報を融合する手法を示す(\ref{subsec:Fused_Embedding}節)
%     \item 公開データセットを用いて実験を行って提案モデルを評価し(\ref{sec:Experiment}節)，
%     グラフ手法の有用性を分析する(\ref{sec:Analysis}節)．
% \end{enumerate}

\section{Related Works}
\label{sec:related-works}
% \subsection{Relationship Classification Between Images and Text}
% \label{subsec:Relation_between_IandT}
% 分類ラベルとして，
% Vempalaらの研究\cite{vempala2019Categorizing}により関係のモデル化が行われている．
% この研究では画像とテキストの両方が存在するツイートに対して,
% テキストが画像内で表現されているかどうかを判別するテキストタスクと，
% 画像がツイートに意味を追加するかどうかを判別する画像タスクの2種類の二値分類を併せて4種類の関係に分類している.
% テキストタスクでは,
% テキストの内容と画像の間に意味的な重複があるかを考慮するために,
% テキストが画像内で表現されているかどうかを判別する.
% 画像タスクでは,
% 画像がテキストの内容を超えてツイートに意味を付加しているかどうかを判別するために,
% 画像がツイートに意味を追加するかどうかを判別する.
% そして人為的にこれら4種類にラベル付けがされた投稿データセットを構築し,
% InceptionNet等の手法を用いて自動分類を行っている．
% 結果として,
% LSTM+InceptionNetのアンサンブルモデルを使用することで画像タスクで優れた結果を示し，
% 画像が冗長かつツイート内容に意味を与えていない場合に対して，特に分類性能が高いことを示している．
% そしてこの研究に対して,
% 後にSunらが上記のVempalaらの研究の投稿データセットについて,
% 人為でラベルをつけたことによる誤ラベリングの存在に着目しリラベリングを行っている\cite{sun2023Unsupervised}．
% またこの研究では,
% 人為でラベル付けする問題に対処するため,
% クラスタリングによって擬似ラベルを生成する教師なし学習手法を用いて,
% 分類器とエンコーダの学習を行なって分類し，
% 性能が改善されている．

% Ottoらは,
% 言語学やコミュニケーション科学とマルチメディアとコンピュータビジョンの研究の隔たりを埋めるため,
% クロスモーダル相互情報,意味相関,および状態関係の三つの次元に基づいて,
% 8種類の意味的画像-テキストクラスの新しい分類を行っている\cite{otto2020Characterization}.
% Xuらは,
% 新たに``Insertion'', ``Concretization'', ``Projection'', ``Restatement'', ``Extension''という5種類に分類し,
% 画像，テキストならびにキャプションのデータに対してMulti-Head Attentionを利用したモデルにより自動分類を行っている\cite{xu2022Understanding}.
% 他にもSoseaらの研究\cite{sosea2021Using}では，
% 災害のマルチモーダルなデータを構築して``Unrelated''，``Similar''，``Complementary''の3種類に分類し，
% 画像とテキストの結合表現を学習するBERTモデル(ViLBERT\cite{lu2019ViLBERTa})等を利用して分類している．
% このように様々な分類軸や技術が提案されている.
Previous studies on image-text relationship classification in SNS have explored various approaches.
Vempala et al. \cite{vempala2019Categorizing} modeled the interactions between images and text and proposed a dataset that categorizes relationships into four types, along with a classification task. 
Their approach combines two binary classification tasks: one determining whether the text is explicitly represented in the image and another assessing whether the image adds meaning to the tweet.
Sun et al.\cite{sun2023Unsupervised} later refined this dataset by correcting mislabeling issues and improving classification performance using unsupervised clustering techniques.
Other classification approaches have also been explored. 
Otto et al.\cite{otto2020Characterization} categorized relationships into eight types based on cross-modal semantic relationships, 
while Xu et al.\cite{xu2022Understanding} developed a dataset with five categories grounded in human perception, 
focusing on entities and scenes.
Additionally, 
Sosea et al.\cite{sosea2021Using} focused on disaster-related data, classifying relationships into three categories, as shown in Figure \ref{fig:fig_relatedexample}. 
They employed multimodal models, such as ViLBERT \cite{lu2019ViLBERTa}, 
a BERT-based model designed to learn the interactions between images and text.
% SNSの画像ーテキスト対の関連性分類に関する先行研究では，
% まずVempalaらが画像とテキストの相互関係をモデル化し，
% 4種類の関係に分類するデータセットの構築とタスクの提案をしている\cite{vempala2019Categorizing}．
% テキストが画像内で表現されているかを判別するテキストタスクと，
% 画像がツイートに意味を追加するかを判別する画像タスクの2種類の二値分類を併せて分類している．
% Sunらはこのデータセットの誤ラベリングを訂正し，
% 加えて教師なし学習によるクラスタリングを活用して分類性能を向上させている\cite{sun2023Unsupervised}．
% また別の関連性の分類法として
% Ottoらはクロスモーダルな意味関係に基づき8種類\cite{otto2020Characterization}，
% Xuらは人間の知覚をふまえてエンティティやシーンに基づいた5種類\cite{xu2022Understanding}の分類に基づくデータセットを構築して分類している．
% またSoseaらは災害データを対象に，
% 図\ref{fig:fig_relatedexample}の3種類に分類を行い，
% 画像と言語の相互関係を学習するBERTモデルであるViLBERT\cite{lu2019ViLBERTa}等のマルチモーダルモデルを用いた分類を実施している\cite{sosea2021Using}．
In this study, 
unlike previous research, 
we propose a classification method that maps image and text data into a shared latent feature space and constructs a graph using clustering. 
Specifically, 
we aim to improve the classification accuracy of complementary relationships by utilizing the dataset \cite{sosea2021Using}.
For edge classification, 
Aggarwal et al.\cite{aggarwal2016Edge} formulated edge classification as a problem solvable via heuristic methods using the Jaccard coefficient. 
Wang et al.\cite{wang2020Edge2vec} introduced Edge2Vec, 
which maps edge information into a low-dimensional space. 
Cheng et al.\cite{cheng2024Edge} proposed a method that addresses the issue of topological imbalance.
In this study, 
we update edge representations using GCN by adopting an approach that reconstructs edges as nodes, 
enabling effective learning of edge representations.
% また，
% グラフ構造を活用したノード分類およびエッジ分類に関する先行研究として，
% Zhuangらは局所的および大域的な情報を考慮したGCNによる半教師あり学習手法\cite{zhuang2018Dual}，
% Liらはラベル共起パターンを考慮したGCNを活用したテキストのマルチラベル分類手法 \cite{li2024Dualview}を提案している．
% エッジ分類では，
% Aggarwalらがエッジ分類を定式化してJaccard係数を用いたヒューリスティック手法\cite{aggarwal2016Edge}，
% Wangらがエッジ情報を低次元空間にマッピングするEdge2Vec\cite{wang2020Edge2vec}，
% Chengらがトポロジーの不均衡問題に着目した手法 \cite{cheng2024Edge} を提案している．
% 本研究では，
% GCN を用いたエッジ表現の更新を行うため，
% エッジをノードとして再構成するアプローチを採用する．

\section{Proposed Method}
\label{sec:method}
\begin{figure}[tb]
    \centering
    \includegraphics[width=\linewidth]{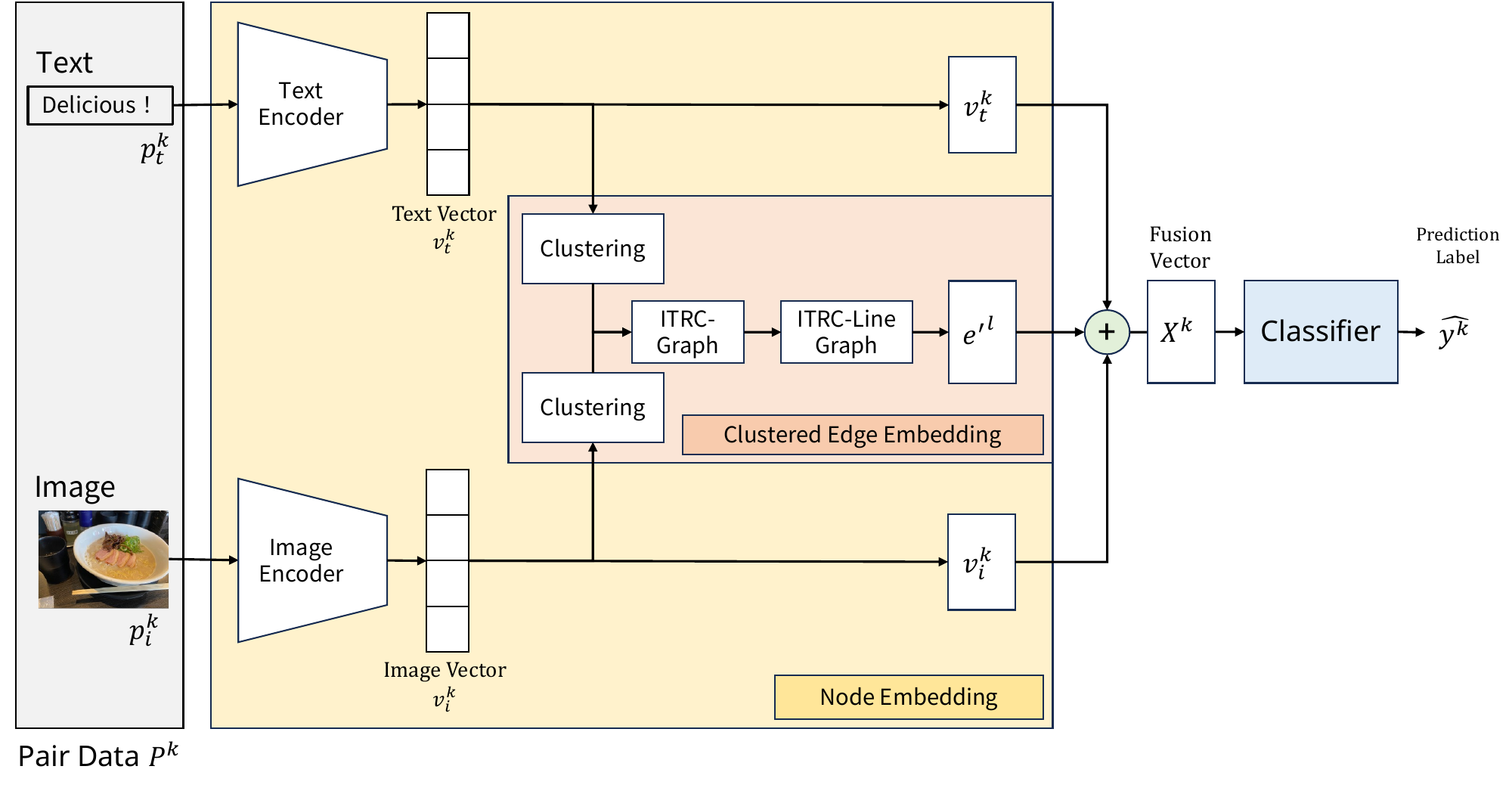}
    \caption{Overview of Proposed Method}
    \label{fig:proposed}
\end{figure}

\subsection{Problem Definition}
\label{subsec:Problem}
For a set of $N$ paired posts $P$, 
the $n$-th data sample consists of an image $i$ and a text $t$, 
represented as a pair $P_n = \{ p^t_n, p^i_n \} $. 
The output of the image-text relationship analysis is the predicted label $\hat{y_n}$, 
which represents the estimated relationship between $p^t_n$ and $p^i_n$. 
The task is to predict one of $C$ relationship types for each $P_n$.
% $N$ペアの全投稿データ$P$に対し，
% $n$番目のデータは画像$i$とテキスト$t$のペア$P_n=\{p^t_n, p^i_n\}$とする.
% 画像ーテキスト対の関連性分析の出力は$p^t_n$と$p^i_n$の関連性の予測ラベル$\hat{y_n}$で,
% $P_n$について$C$種類の関連性から1種類を予測する.

\subsection{Overview of the Proposed Method}
\label{subsec:Proposed_Method_abs}
Figure 2 illustrates the overall workflow of the proposed method.
% Based on Figure \ref{fig:proposed}, 
% we describe the overall workflow of the proposed method.
% 図\ref{fig:proposed}に基づき，
% 本手法の全体の流れを説明する．

First, 
in the Node Embedding stage, 
the image-text pair data is embedded to obtain the respective representation vectors $\mathbf{v}^t_n$ and $\mathbf{v}^i_n$ (Section \ref{subsec:Node_Embedding}).
% まず，
% Node Embedding部では，
% 画像とテキストのペアデータをエンべディングし，
% 各表現ベクトル($\mathbf{v}^t_n$, $\mathbf{v}^i_n$)を得る(\ref{subsec:Node_Embedding}節)．

Next, 
in the Clustered Edge Embedding stage, 
two major processes are performed.
First, 
the text and image embeddings are clustered to construct the ITRC-Graph $G'$, 
where each cluster serves as a node (Section \ref{subsubsec:ITRC-Graph}).
Then, 
$G'$ is transformed into the ITRC-Line Graph $G^*$, 
where nodes and edges are interchanged.
Through representation learning using GCN, 
the node embeddings of $G^*$ are updated, 
yielding edge embeddings for $G'$ (Section \ref{subsubsec:ITRC-Line-Graph}).
To prevent information loss, 
the intermediate layer embeddings of the $G^*$ learning model are utilized as edge vectors $\mathbf{v}^e_n$ (Section \ref{subsec:Clustered Edge Embedding}).
% 次に，
% Clustered Edge Embedding部では，
% 大きく二つの処理を行う．
% まず，
% テキストおよび画像のエンベディングをクラスタリングし，
% 各クラスタをノードとしたITRC-Graph $G'$ を構築する(\ref{subsubsec:ITRC-Graph}節)．
% 次に$G'$に対してエッジとノードを入れ替えたITRC-Line Graph $G^*$ に変換し，
% GCNによる表現学習を通じて$G^*$のノードのエンべディングを更新して
% $G'$ のエッジエンべディングを得る(\ref{subsubsec:ITRC-Line-Graph}節)．
% 本手法では情報の損失を防ぐため，
% $G^*$の学習モデルの中間層のエンべディングをエッジベクトル $\mathbf{v}^e_n$ として活用する(\ref{subsec:Clustered Edge Embedding}節)．

Finally, 
in the fusion stage,
the text embedding, 
the image embedding, 
and the edge embedding \( (\mathbf{v}^t_n, \mathbf{v}^i_n, \mathbf{v}^e_n) \) are fused (Sections \ref{subsec:Fused_Embedding}).
Then,
in the classification stage, 
the fusion vector is classified using an MLP model (Section \ref{subsec:Categorization}).
% 最後に，
% 融合・分類のステップでは，
% テキスト，画像，エッジのエンべディング ($\mathbf{v}^t_n$, $\mathbf{v}^i_n$, $\mathbf{v}^e_n$) を統合し，
% MLPモデルにより分類を行う(\ref{subsec:Fused_Embedding}, \ref{subsec:Categorization}節)．

\subsection{Image and Text Encoding}
\label{subsec:Node_Embedding}
For the entire set of SNS posts $P$, 
the image-text pair data \( P_n=\{p^t_n, p^i_n\} \) is embedded using an encoder, 
transforming the text and image into 512-dimensional vectors \( \mathbf{v}^t_n \) and \( \mathbf{v}^i_n \), 
respectively. 
Then, 
for each pair, 
a vector set $V_n =\{\mathbf{v}^t_n, \mathbf{v}^i_n\}$ is constructed, 
and the entire dataset is represented as the set of all pairs' vectors $V=\{V_1, V_2, \dotsc, V_N\}$.
% SNS上の全投稿データ$P$に対し，
% 画像とテキストのペアデータ$P_n=\{p^t_n, p^i_n\}$をエンコーダでエンべディングし，
% テキストと画像をそれぞれ512次元のベクトル$\mathbf{v}^t_n$および$\mathbf{v}^i_n$に変換する．
% そしてペア毎のベクトル集合$V_n(=\{\mathbf{v}^t_n, \mathbf{v}^i_n\})$を構築し，
% 全ペアのベクトル集合$V(=\{V_1, V_2, \dotsc, V_N\})$を得る．  
Here, 
the encoder employed is CLIP \cite{radford2021Learning}.
% a multimodal model capable of mapping images and text into a shared latent feature space. 
% CLIP is trained such that the cosine similarity between paired images and texts is maximized. 
% Leveraging this property, 
% we aim to achieve embeddings that effectively capture the relationships between images and text.
% ここでエンコーダには，
% 画像とテキストを共通の潜在特徴空間にマッピング可能なマルチモーダルモデルであるCLIP\cite{radford2021Learning}を用いる．
% CLIPは，
% 画像とテキストのペアのコサイン類似度が高くなるように学習されており，
% この特性を活用して関連情報を考慮したエンべディングを実現することを期待する．

\subsection{Embedding of Clustered Pseudo-Graphs}
\label{subsec:Clustered Edge Embedding}
\begin{figure*}[t]
    \centering
    \includegraphics[width=\linewidth]{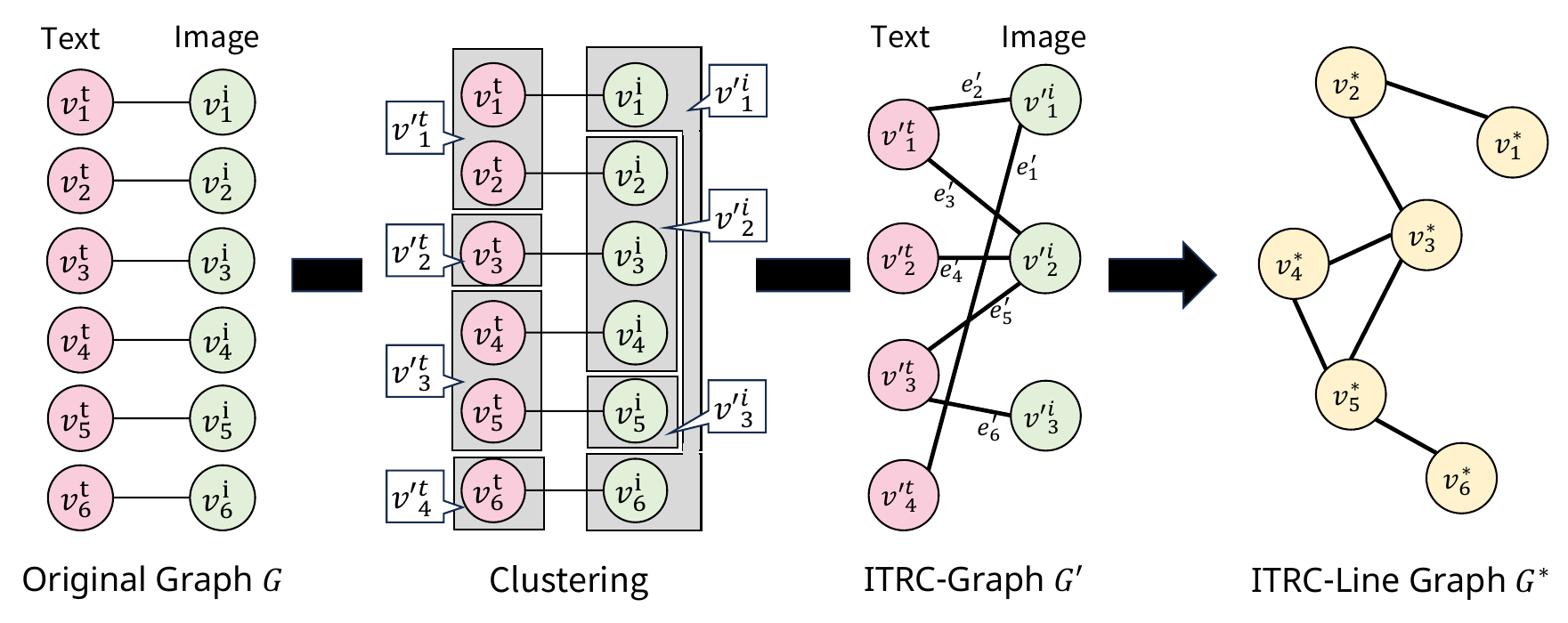}
    \caption{The process  of Clustered Edge Embedding}
    \label{fig:fig_Clustered_Edge_Embedding}
\end{figure*}

The model's input consists of the embedding set of image-text pairs $V$, 
and the output is the node embeddings in the ITRC-Line Graph $G^*$, 
denoted as $\{\mathbf{v}^*_1, \mathbf{v}^*_2, \dotsc, \mathbf{v}^*_L\}$, 
where $L$ represents the number of nodes in $G^*$.
% 入力は画像-テキストペアのエンべディング集合$V$であり，  
% 出力はITRC-Line Graph $G^*$におけるノードのエンべディング$\{\mathbf{v}^*_1, \mathbf{v}^*_2, \dotsc, \mathbf{v}^*_L\}$である
% ($L$は$G^*$のノード数)．  

\subsubsection{Construction of ITRC-Graph}
\label{subsubsec:ITRC-Graph}
First, 
for the SNS post dataset $P$, 
a heterogeneous graph $G$ is constructed, 
where each image and text is treated as a node, 
and an edge is formed between them in the paired data. 
While this edge representation captures relationships within pairs, 
the graph structure is sparse, 
making it challenging for GNN-based learning to leverage information across pairs.
% まずSNSの投稿データ$P$に対し，
% 各画像とテキストをノードとし，
% ペアデータ間にエッジがあるヘテログラフ$G$を構築する．
% これのエッジ表現によりペア内部の関係を捉えることが可能になると考えられるが，
% スパースな構造であり，
% ペア間の情報を活用するためのGNNによる学習が困難である問題がある．

To address this issue, 
clustering is applied, 
as illustrated in Figure \ref{fig:fig_Clustered_Edge_Embedding}, 
to construct the ITRC-Graph $G'=\{\mathcal{V}', \mathcal{E}'\}$, 
where each cluster is treated as a node. 
This approach transforms the sparse structure into a denser graph, 
facilitating information propagation across pairs. 
The construction procedure is as follows:
% この問題を解決するため，
% 図\ref{fig:fig_Clustered_Edge_Embedding}のようにクラスタリングを適用し，
% 各クラスターをノードとするITRC-Graph $G'=\{\mathcal{V}', \mathcal{E}'\}$を構築することで，
% 密なグラフ構造を実現しつつ，
% ペア間の情報を活用する．
% 具体的な構築手順は以下の通りである．

\begin{enumerate}
    \item \textbf{Applying K-means Clustering} \\
    The number of clusters $K$ is set, 
    and K-means clustering is applied separately to all images and texts in the entire pair set $V$, 
    dividing them into $K$ clusters each (a total of $2K$ clusters).
    \label{ITCRG_step1}
    
    \item \textbf{Defining the Node Set $\mathcal{V}'$ of the ITRC-Graph $G'$ } \\
    Each cluster is treated as a pseudo-node, 
    forming the node set $\mathcal{V}'$. 
    The embedding of each node is computed as the mean vector of the embeddings belonging to the corresponding cluster.
    \label{ITCRG_step2}

    \item \textbf{Defining the Edge Set \( \mathcal{E}' \) of the ITRC-Graph \( G' \)} \\
    The edge set $\mathcal{E}'=\{{e'}_1, {e'}_2, \dotsc, {e'}_L\}$ is constructed by connecting clusters to which the pairs $P_n$ belong. 
    Each edge $e'_l$ connects a node from a text cluster and a node from an image cluster, 
    ensuring that no duplicate edges exist between the same pair of nodes. 
    The embedding of each edge $\mathbf{e'}_l$ is initialized by concatenating the corresponding two vectors, 
    resulting in a 1024-dimensional representation.
    \label{ITCRG_step3}

    \item \textbf{Assigning Labels to Edges in the ITRC-Graph \( G' \)} \\
    To facilitate GCN learning,
    a single label is assigned to each edge. 
    In cases where multiple labels exist, 
    the majority vote determines the final label. 
    For example, if an edge $(v'^{t}_1, v'^{i}_1)$ is fused with three original pair data points with labels $\{$"$A$", "$B$", "$B$"$\}$, 
    the label "$B$" is assigned to that edge.
    \label{ITCRG_step4}
\end{enumerate}

% \begin{enumerate}
%     \item K-means クラスタリングの適用 \\
%     クラスター数$K$を設定し，
%     全ペア集合$V$の画像とテキスト全てに対して別々にK-meansクラスタリングを適用することで各$K$個のクラスターに分割する(計$2K$個)．
%     \label{ITCRG_step1}
    
%     \item ITRC-Graph $G'$ のノード集合$\mathcal{V}'$の定義 \\
%     各クラスターを擬似ノードとして，
%     そのノード集合を$\mathcal{V}'$とする．
%     この時，
%     各ノードのエンべディングは各クラスターに属するエンべディングの平均ベクトルを計算する．
%     % 画像クラスターのノード集合を$\mathcal{V}'^i=\{{v'}^i_1, {v'}^i_2, ..., {v'}^i_K\}$，テキストクラスターのノード集合を$\mathcal{V}'^t=\{{v'}^t_1, {v'}^t_2, ..., {v'}^t_K\}$とし，統合して$\mathcal{V}' = \{\mathcal{V}'^t, \mathcal{V}'^i \}$を構築する．
%     \label{ITCRG_step2}

%     \item ITRC-Graph $G'$ のエッジ集合$\mathcal{E}'$の定義 \\
%     各ペアデータ$P_n$が属するクラスター同士を接続するエッジ集合
%     $\mathcal{E}'=\{{e'}_1, {e'}_2, \\ \dotsc, {e'}_L\}$を構成する．
%     各${e'}_l$の両端の一方はテキストクラスターによるノード，
%     一方は画像クラスターによるノードとなっており，
%     同じ二つのノード間に二重にエッジが構成されないようにする．
%     % エッジ${e'}_l = \{{v'}^t_a, {v'}^i_b\}$は，テキストクラスター${v'}^t_a$と画像クラスター${v'}^i_b$の間に張られ，二重にエッジが構成されることを防ぐ．
%     また各エッジのエンベディング$\mathbf{e'}_l$は，
%     対応する2つのベクトルを単純に連結して1024次元にすることで初期化される．
%     \label{ITCRG_step3}

%     \item ITRC-Graph $G'$ の各エッジのラベル付与 \\
%     GCNの学習に用いるため，
%     各エッジに対して1種類のラベルを付与する．
%     複数のラベルが存在する場合は，
%     多数決により決定する．
%     例えば，エッジ$(v'^{t}_1, v'^{i}_1)$に対し，
%     属する元のペアデータが3個あり，
%     各ラベルが$\{``A'', ``B'', ``B''\}$である場合，
%     そのエッジにはラベル$``B''$が付与される．
%     \label{ITCRG_step4}
% \end{enumerate}

\subsubsection{ITRC-Line Graph Construction and Representation Learning}
\label{subsubsec:ITRC-Line-Graph}

\begin{figure}[t]
    \centering
    \includegraphics[width=0.9\linewidth]{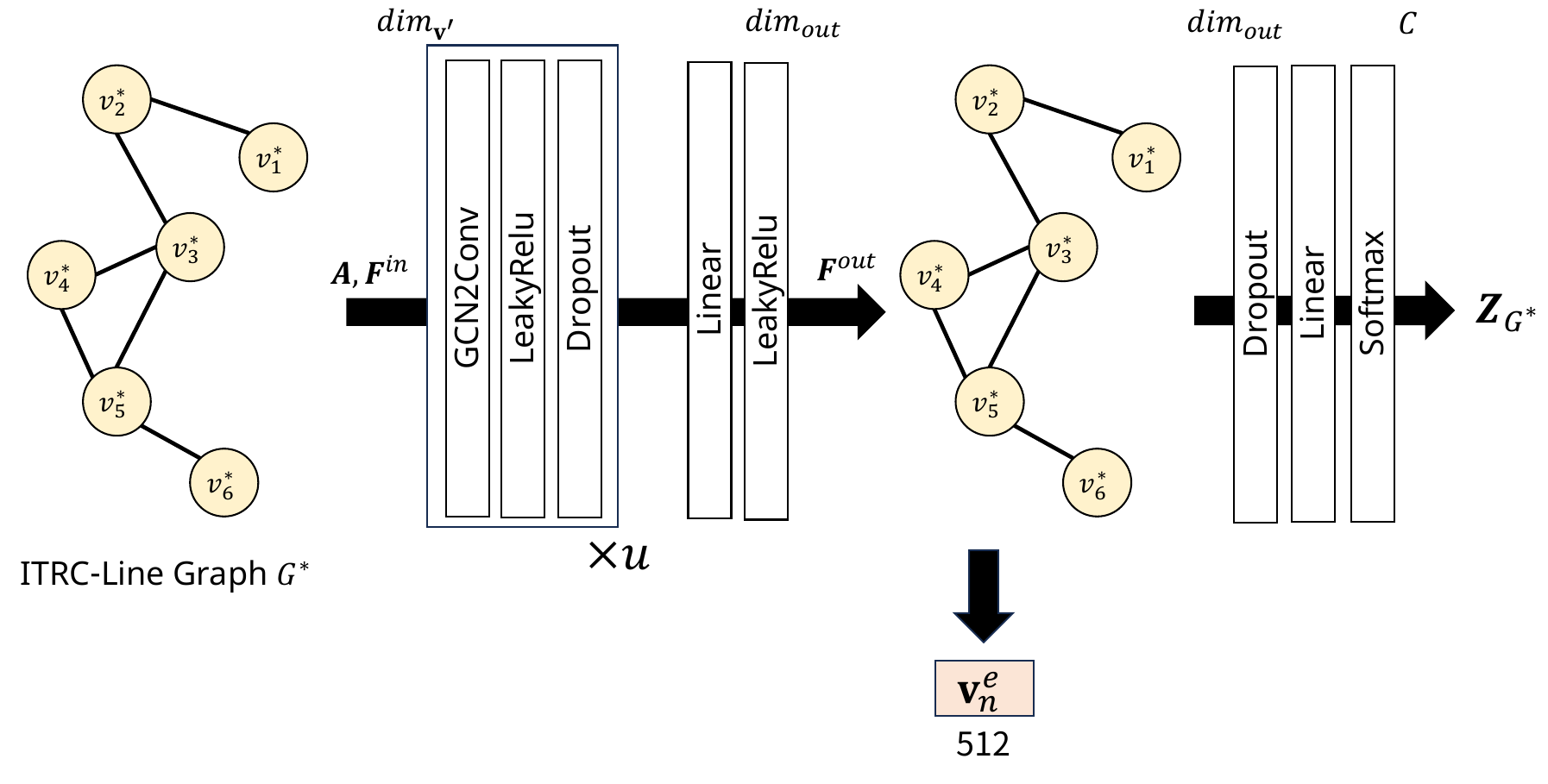}
    \caption{Learning Model of ITRC-Line Graph}
    \label{fig:fig_GstarGraphLearning}
\end{figure}
For the ITRC-Graph $G'$ constructed in the previous section, 
learning is performed using the labels assigned in Step \ref{ITCRG_step4}.
However, 
to apply GCN, 
edges must be treated as nodes. 
To achieve this, 
the ITRC-Line Graph $G^*=\{\mathcal{V}^*, \mathcal{E}^*\}$ is constructed by swapping the nodes and edges of \( G' \), 
as shown in Figure \ref{fig:fig_Clustered_Edge_Embedding}. 
Here, 
$\mathcal{V}^*=\{v^*_1, v^*_2, \dotsc, v^*_L\}$, 
where each node $v^*_l$ corresponds to an edge $e'_l$ in $G'$.
The edge set $\mathcal{E}^*$ is defined by constructing edges between all pairs of nodes in $G^*$ that share an endpoint in $G'$.

% 前節で構築した ITRC-Graph $G'$ に対し，
% Step\ref{ITCRG_step4}で付与されたラベルを用いて学習を行う．
% しかし，
% GCN を適用するためにはエッジをノードとして扱う必要がある．
% そこで，
% 図\ref{fig:fig_Clustered_Edge_Embedding}のように
% ITRC-Graph $G'$ のノードとエッジを入れ替えた ITRC-Line Graph $G^*=\{\mathcal{V}^*, \mathcal{E}^*\}$ を構築する．
% ここで$\mathcal{V}^*=\{v^*_1, v^*_2, \dotsc, v^*_L\}$であり，
% 各ノード $v^*_l$ は $G'$ のエッジ $e'_l$ に対応する．
% またエッジ集合$\mathcal{E}^*$は，
% $G'$の各エッジの両端のノードを共有する全ペア間にエッジが構築され，
% その集合である．

To update the edge representations in ITRC-Graph $G'$, 
a model based on GCNII \cite{chen2020Simple} is constructed, 
as illustrated in Figure \ref{fig:fig_GstarGraphLearning}. 
Using this GCN model, 
supervised learning is performed to learn node representations in the ITRC-Line Graph $G^*$. 

The model takes as input the initial node embedding matrix:

\begin{equation}
\mathbf{F}^{in} = \begin{bmatrix} \mathbf{v}^*_1 & \cdots & \mathbf{v}^*_L\end{bmatrix}^\top \in \mathbb{R}^{L \times dim_{v'}}
\end{equation}
and applies $u$ layers of GCNII convolutional layers (GCN2Conv), 
LeakyReLU, 
and Dropout. 
A fully connected layer then transforms the embeddings into $dim_{out} = 512$ dimensions, 
followed by a LeakyReLU activation. 
The resulting intermediate embedding matrix is denoted as:
$ \mathbf{F}^{out} \in \mathbb{R}^{L \times dim_{out}} $.
Subsequently, 
Dropout is applied to \( \mathbf{F}^{out} \), 
followed by another fully connected layer that maps the embeddings to $C$ dimensions. 
Finally, 
the class probabilities of each node in $G^*$ are obtained via the Softmax function:

\begin{equation}
\mathbf{F}^{out} = f_{GCNmodel}( \mathbf{A}, \mathbf{F}^{in}; \mathbf{\Theta_1}) 
\label{eq:ITRC-Line_1}
\end{equation}

\begin{equation}
\mathbf{Z}_{G^*}= \text{softmax}(\mathbf{F}^{out} \mathbf{\Theta_2}) 
\label{eq:ITRC-Line_2}
\end{equation}
where $\mathbf{A} \in \mathbb{R}^{L \times L}$ is the adjacency matrix of \( G^* \),  
$f_{GCNmodel}(\cdot)$ represents the GCN model,  
$\mathbf{\Theta_1}$ denotes its trainable parameters, and  
$\mathbf{\Theta_2} \in \mathbb{R}^{dim_{out} \times C}$ represents the parameters of the output layer.
% ITRC-Graph $G'$ のエッジ表現を更新するために，
% \ref{subsec:rw_graph}節で言及したGCNII \cite{chen2020Simple} を用いた図\ref{fig:fig_GstarGraphLearning}のようなモデルを構築し，
% そのGCNモデルを用いた教師あり学習により，
% ITRC-Line Graph $G^*$ の各ノードの表現学習を行う．
% モデルは，
% 初期ノードエンベディング行列の$\mathbf{F}^{in} = \begin{bmatrix} \mathbf{v}^*_1 & \cdots & \mathbf{v}^*_L\end{bmatrix}^\top \in \mathbb{R}^{L \times dim_{v'}}$を入力として，
% GCNIIの畳み込み層(GCN2Conv), LeakyReLU, Dropoutを$u$回重ね，
% 1層の全結合層で$dim_{out}=512$次元に線形変換して，
% LeakyReLUを適用する．
% ここでの中間層のエンべディング行列を$\mathbf{F}^{out}\in \mathbb{R}^{L \times dim_{out}}$とする．
% 更に，
% $\mathbf{F}^{out}$についてDropoutを行い，
% 1層の全結合層によって$C$次元へ変換し，
% Softmaxにより，
% 各ノードのクラス確率$\mathbf{Z}_{G^*} \in \mathbb{R}^{L \times C}$を出力する．
% 以上をまとめると，
% 以下の式で定義される：
% \begin{equation}
%     \mathbf{F}^{out} = f_{GCNmodel}( \mathbf{A}, \mathbf{F}^{in}; \mathbf{\Theta_1}) 
%     \label{eq:ITRC-Line_1}
% \end{equation}
% \begin{equation}
%     \mathbf{Z}_{G^*}= \text{softmax}(\mathbf{F}^{out} \mathbf{\Theta_2}) \label{eq:ITRC-Line_2}
% \end{equation}
% ここで，
% $\mathbf{A} \in \mathbb{R}^{L \times L}$ は $G^*$ の隣接行列，  
% % $\mathbf{F}^{in} \in \mathbb{R}^{L \times dim_{v'}}$ は初期エンベディング，  
% $f_{GCNmodel}(\cdot)$ は GCN モデル，
% $\mathbf{\Theta_1}$ はその学習パラメータ，
% $\mathbf{\Theta_2} \in \mathbb{R}^{dim_{out} \times C}$ は出力層のパラメータである．

Since the dimension of $\mathbf{Z}_{G^*}$ is constrained by the number of classes $C$,
the learned intermediate embedding $\mathbf{F}^{out}$ is extracted after training. 
Specifically, 
the embedding of the node $v^*_l$ corresponding to the edge between the clusters of the original pair $P_n$ is used as the edge embedding \( \mathbf{v}^e_n \), yielding a 512-dimensional representation.
% なお，
% $\mathbf{Z}_{G^*}$ の次元数はクラス数 $C$ に制約される．
% そこで学習後の$\mathbf{F}^{out}$を取り出し，
% 元のペア$P_n$のクラスター間のエッジが属する$v^*_l$のエンべディングを$\mathbf{v}^e_n$として，
% 512 次元のエンベディングを得る．

\subsection{Fusion of Representations}
\label{subsec:Fused_Embedding}
Through the Node Embedding and Clustered Edge Embedding processes,  
we obtain three types of 512-dimensional vectors:  
the text vector $\mathbf{v}^t_n$,  
the image vector $\mathbf{v}^i_n$,  
and the node vector of the ITRC-Line Graph  
(i.e., the edge vector of the ITRC-Graph) $\mathbf{v}^e_n$.  
These vectors are fused to form the final fused vector $\mathbf{X}_n$  
using one of three fusion methods.  
The proposed fusion methods are listed as follows:  
\begin{itemize}
    \item \textbf{Method 1 (Averaging)}: Preserves the original dimensionality by calculating the mean of the vectors.  
    \item \textbf{Method 2 (Concatenation)}: Increases the dimensionality while retaining all information by stacking the vectors.  
    \item \textbf{Method 3 (Averaging + Concatenation)}: Takes an intermediate approach between Method 1 and Method 2 by averaging the text and image vectors, then concatenating the result with the edge vector.  
\end{itemize}
These fusion methods combine all three vectors before classification.

\subsection{Classification of Fusion Representations}
\label{subsec:Categorization}
The classification is performed using a three-layer MLP, 
taking the fused vector $\mathbf{X}_n$ generated in Section \ref{subsec:Fused_Embedding} as input.  
After the linear layers in the first and second layers, ReLU activation and dropout are applied to prevent overfitting.  
Finally, 
the predicted label $\hat{y^n}$ is output using Softmax.  

The input to the classifier has the same dimensionality as $\mathbf{X}_n$,  
while the output is a $C$-dimensional vector corresponding to the number of labels.  
The class with the highest likelihood is selected as the predicted label $\hat{y^n}$.

% \ref{subsec:Fused_Embedding} 節で生成した融合ベクトル$\mathbf{X}_n$を入力とし，
% 3層MLPにより分類を行う．  
% 1層目と2層目の線形層の後にReLUによる活性化とDropoutによる過学習抑制を行う．
% 最終的にSoftmaxにより予測ラベル $\hat{y^n}$ を出力する．

% 分類器の入力は $\mathbf{X}_n$ と同じ次元数で，  
% 出力はラベル数 $C$ に対応する $C$ 次元のベクトルとなる．  
% 最も尤度の高いクラスが予測ラベル $\hat{y^n}$ となる．

\section{Experiment}
\label{sec:Experiment}
We compare multiple models proposed in Section \ref{sec:method} 
through experiments with various variations and discuss the optimal model.  
% \ref{sec:method}節で提案した複数のモデルについて，
% 様々なバリエーションによる実験で比較し，
% 最適なモデルについて議論する．

\subsection{Experiment Setup}
\label{secsub:Setup}
\subsubsection{Dataset}
\label{secsubsub:Dataset}
In this study, 
we use DisRel \cite{sosea2021Using}.
% as described in Section \ref{sec:intro}.  
The distribution of the three label types is as follows:  
``Complementary'' consists of 1,781 pairs,  
``Similar'' consists of 2,919 pairs,  
and ``Unrelated'' consists of 291 pairs,  
resulting in a total of 4,991 pairs.  
Since the previous study \cite{sosea2021Using} excluded ``Unrelated'' from the analysis,  
we also conducted our experiments using only the two classes, 
``Complementary'' and ``Similar''.  
The number of data pairs used is $N=4700$\footnote{Although the paper \cite{sosea2021Using} states 4,600 pairs,
the dataset has been updated to version 2 as of March 23, 2025,
containing 4,700 pairs (\url{https://github.com/tsosea2/DisRel}).},  
and the number of classification classes is $C=2$.
% 本研究では，
% \ref{sec:intro}節で述べたDisRel\cite{sosea2021Using}  
% を使用する．  
% 各3種類のラベルの分布は，
% ``Complementary''が1781ペア，
% ``Similar''が2919ペア，
% ``Unrelated''が291ペアで計4991ペアである．
% 先行研究\cite{sosea2021Using}にて，
% ``Unrelated''を分析対象から除外していることから，
% 本研究の実験においても，
% ``Complementary''と``Similar''の 2クラスで実験を行った．  
% 使用したデータ数は$N=4700$ペア\footnote{論文\cite{sosea2021Using}には4600ペアとあるが, \today 現在v2に更新されており4700ペアある(\url{https://github.com/tsosea2/DisRel})}
% であり，  
% 分類クラス数は$C=2$である．

\subsubsection{Analysed Models}
\label{subsubsec:Compare_Originalmodel}
\begin{table}[t]
    \centering
    \caption{Comparison of fused vector models: ``T'' represents Text, ``I'' represents Image, and ``E'' represents Edge. ``A'' represents average method, and ``C'' represents concatenation method.}
    \label{tab:Fused_Vector_Models}
    \begin{tabular}{cccccc}
        \hline
        Model Name & Model Type     & T & I & E & Fusion Method \\
        \hline
        T+I(A) & Baseline  & \checkmark & \checkmark &           & Average \\
        T+I(C) & Baseline  & \checkmark & \checkmark &           & Concatenation \\
        T+E(A) & Proposed  & \checkmark &            & \checkmark & Average \\
        T+E(C) & Proposed  & \checkmark &            & \checkmark & Concatenation \\
        I+E(A) & Proposed  &            & \checkmark & \checkmark & Average \\
        I+E(C) & Proposed  &            & \checkmark & \checkmark & Concatenation \\
        T+I+E(A) & Proposed  & \checkmark & \checkmark & \checkmark & Average \\
        T+I+E(C) & Proposed  & \checkmark & \checkmark & \checkmark & Concatenation \\
        T+I+E(A+C) & Proposed  & \checkmark & \checkmark & \checkmark & Average + Concatenation \\
        \hline
    \end{tabular}
\end{table}
As summarized in Table~\ref{tab:Fused_Vector_Models},
we construct nine comparative models based on the fusion of three vectors ($\mathbf{v}^t_n$, $\mathbf{v}^i_n$, $\mathbf{v}^e_n$).  
Six models fuse two vectors using averaging or concatenation, 
while three models combine all three vectors using averaging, concatenation, or both.  
The two models without edge vectors are baselines, 
while the remaining seven models with edge vectors are proposed models.

% Based on the composition of the fused vector $\mathbf{X}_n$, 
% we construct nine comparative models.  
% Six models fuse any two of the three vectors ($\mathbf{v}^t_n$, $\mathbf{v}^i_n$, $\mathbf{v}^e_n$) using averaging (Method 1) or concatenation (Method 2).  
% Three additional models combine all three vectors using averaging (Method 1), concatenation (Method 2), or average + concatenation (Method 3).  
% The two models without edge vectors serve as baselines, 
% while the remaining seven models including edge vectors are treated as proposed models.  

In the experiments, 
we compare these models to assess the effective utilization of multimodal information and evaluate the proposed methods against existing models \cite{sosea2021Using}.  
Additionally,
we also compare models that use Edge2Vec\cite{wang2020Edge2vec} to directly embed the edges of $G'$.
Since the edge vectors embed by Edge2Vec have 64 dimensions, they are concatenated to preserve information.

% Considering the composition of the fused vector $\mathbf{X}_n$ and the differences in fusion methods, we create nine comparative models.  
% Specifically, six models are designed by fusing any two of the three vectors ($\mathbf{v}^t_n$, $\mathbf{v}^i_n$, $\mathbf{v}^e_n$) using either averaging (Method 1) or concatenation (Method 2).  
% Additionally, three models are constructed by fusing all three vectors ($\mathbf{v}^t_n$, $\mathbf{v}^i_n$, $\mathbf{v}^e_n$) using averaging (Method 1), concatenation (Method 2), or average + concatenation (Method 3).  
% Among these models, two that do not use edge vectors serve as baselines, while the remaining seven models that include edge vectors are treated as proposed models.  

% In the experiments, we compare these models to examine effective utilization of diverse information and evaluate the proposed methods by comparing them with existing models \cite{sosea2021Using}.

% 融合ベクトル $\mathbf{X}_n$ の構成と融合手法の違いを考慮し，
% 9種類の比較モデルを作成する．
% 具体的には，
% ($\mathbf{v}^t_n$, $\mathbf{v}^i_n$, $\mathbf{v}^e_n$)のいずれか2種類のベクトルを平均(方法1)または連結(方法2)で融合する手法(計6種類)，
% もしくは($\mathbf{v}^t_n$, $\mathbf{v}^i_n$, $\mathbf{v}^e_n$)の3種類のベクトルを平均(方法1)または連結(方法2)または平均+連結(方法3)で融合する手法(計3種類)である．
% ここでエッジベクトルを使用しない2種類のモデルをベースラインとし，
% エッジベクトルを含む7種類を提案モデルとする．

% 実験では，
% 各モデルの比較を行い，
% 多様な情報の有効な利用方法を検討するとともに，
% 既存モデル\cite{sosea2021Using} との比較を通じて提案手法の評価を行う．

\subsubsection{Model Implementation}
\label{secsubsub:Model_Setup}

\paragraph{Node Embedding}
For text and image encoding, 
CLIP encoders are used.  
The original study \cite{radford2021Learning} compared ResNet and ViT as image encoders,  
demonstrating that ViT outperforms ResNet.  
Therefore, we adopt the ViT model in this study.  
Specifically, we use the pre-trained CLIP model "ViT-B/32"\footnote{\url{https://github.com/OpenAI/CLIP}},  
from which we obtain 512-dimensional vectors for both text and image representations.  

% テキスト及び画像のエンコードには，
% CLIPのエンコーダを用いる．
% 画像エンコーダは，
% 原著\cite{radford2021Learning}にてResNetとViTを比較しているが，
% ViTの方が性能が高い結果を示していることから，
% ViTモデルを使用する．
% CLIPの事前学習済みモデルの``ViT-B/32''\footnote{\url{https://github.com/OpenAI/CLIP}}を使用し，
% 各512次元のベクトルを得る．

\paragraph{Clustered Edge Embedding}
We apply K-means \cite{lloyd1982Least} clustering with $K=100$ and cosine similarity to classify images and texts into 100 clusters using the "KmeansClusterer" from the nltk.cluster library \cite{bird2009Natural}.
Next, 
we construct the ITRC-Graph $G'$ with clusters as nodes. 
To avoid instability and computational complexity, 
edge reduction is performed by removing edges with the same label between training nodes and connecting only the $J=5$ nearest nodes based on Euclidean distance.

For edge labeling in $G'$, 
majority voting is applied: 
if all edges belong to training data, 
the label is determined by majority voting; 
if both training and test data coexist, 
only training data labels are considered; 
in the case of a tie, 
a label is chosen randomly; 
if all edges belong to test data, 
the label is determined within the test nodes.

Finally, 
we construct the ITRC-Line Graph $G^*$ from $G'$ and train the model using PyTorch Geometric \cite{fey2019Fast}. 
The GCNII layers are set to $u=64$, 
learning rate to 0.005, 
and optimizer to Adam \cite{kingma2017Adam}. 
The loss function is cross-entropy error, 
and training is done for 100 epochs with learning rate decay. 
The evaluation of $G^*$ is discussed in Section \ref{subsec:Analysis_ITRC-LineGraph}.

\paragraph{Fusion and Classification}
We generate the fused vector $\mathbf{X}_n$ for the 9 models described in Section \ref{subsubsec:Compare_Originalmodel}.  
Then, we implement and train a three-layer MLP using PyTorch.  
During training with this model, the batch size is set to 4,  
and the optimizer used is Adam \cite{kingma2017Adam}.  
The loss function is set to cross-entropy error.  
We apply learning rate decay using the LambdaLR scheduler,  
and training is conducted for 100 epochs.  
If the loss does not improve for a certain period, training is stopped at that point.  
The model with the lowest validation loss is selected for evaluation.  

% All training experiments are performed on an Nvidia GPU,  
% specifically the GeForce GTX 1080. 
% \ref{subsubsec:Compare_Originalmodel}節で述べた9種類のモデルについて融合ベクトル$\mathbf{X}_n$を生成する．
% そして，
% 3層MLPを用い，
% PyTorchで実装・学習を行う．
% このモデルでの学習では，
% バッチサイズは4とし，
% オプティマイザはAdam\cite{kingma2017Adam}を使用し，
% 損失関数は交差エントロピー誤差を用いる.
% またLambdaLRのスケジューラにより学習率を減衰させつつ100エポックの学習を行うが，
% 損失の更新が一定進まない場合には学習をそこで止める．
% そして検証データの損失が最も下がった時のモデルを利用する．
% これらの実験の学習は全てNvidia GPUのGeForce GTX 1080を用いた．

\subsubsection{Evaluation}
\label{secsubsub:Evaluation_Method}
In this experiment, we compare the classification performance of the models based on the presence or absence of image, text, and ITRC-Line Graph node vectors,  
as well as the fusion methods described in Section \ref{subsec:Fused_Embedding}.  
Specifically, we evaluate the 9 models proposed in Section \ref{subsubsec:Compare_Originalmodel},  
along with a comparison against existing models.  

For evaluation, the 4,700 pairs from the DisRel dataset are randomly split  
into training, validation, and test sets in a ratio of $6:2:2$.  
The evaluation metrics include precision, recall, and F1 score for each label,  
as well as the macro-F1 score and accuracy, calculated solely on the test data.  

% 本実験では，
% 画像・テキスト・ITRC-Line Graph のノードベクトルの有無，
% および\ref{subsec:Fused_Embedding}節の融合手法別の分類性能，
% すなわち\ref{subsubsec:Compare_Originalmodel}節の9種類のモデルを比較し，
% 更に既存モデルとの比較を行う．
% 評価について，
% 4700個のDisRelデータセットを，
% 無作為に$\text{Train}:\text{Validation}:\text{Test}=6:2:2$に分割する．

% 評価指標については，
% テストデータのみに対するラベル毎の適合率，
% 再現率，
% F1値，
% Macro-F1, 
% および正解率を用いる．

\subsection{Results}
\label{subsec:result}

\begin{table}[t]
    \begin{center}
        \caption{Comparison of proposed models by fusion methods and existing models (rounded to three decimal places, average with various of 10 times)}
        % \caption{融合方法別の提案モデルの比較(小数第3位を四捨五入，10回の平均値)}
        % \vspace{-1pt}
        \begin{tabular}{ccccccccc} \toprule
             & \multicolumn{3}{c}{Similar} & \multicolumn{3}{c}{Complementary} & Macro F1 & Accuracy \\ \cmidrule{2-7}
            & P & R & F1 & P & R & F1 & &\\ \midrule
            \begin{tabular}{c}ViLBERT-REL-MT \\ (Best Model from \cite{sosea2021Using}) \end{tabular} & {\bf 0.82} & {\bf 0.82} & {\bf 0.82} & 0.63 & 0.62 & 0.62 & 0.72 & {\bf 0.76} \\ 
            T+E(Edge2Vec \cite{wang2020Edge2vec}) & 0.79 & {\bf 0.82} & 0.80 & {\bf 0.69} & 0.65 & {\bf 0.67} & {\bf 0.74} & 0.75 \\
            I+E(Edge2Vec \cite{wang2020Edge2vec}) & 0.72 & {\bf 0.82} & 0.76 & 0.62 & 0.47 & 0.53 & 0.65 & 0.69 \\
            T+I(A)+E(Edge2Vec \cite{wang2020Edge2vec}) & 0.78 & {\bf 0.82} & 0.80 & 0.68 & 0.63 & 0.66 & 0.73 & 0.75 \\
            T+I(C)+E(Edge2Vec \cite{wang2020Edge2vec}) & 0.78 & {\bf 0.82} & 0.80 & {\bf 0.69} & 0.63 & 0.66 & 0.73 & 0.75 \\ \midrule
            T+I(A) & 0.78 & {\bf 0.82} & 0.80 & 0.68 & 0.63 & 0.65 & 0.72 & 0.74 \\ %T
            T+I(C) & 0.78 & {\bf 0.82} & 0.80 & {\bf 0.69} & 0.63 & 0.66 & 0.73 & 0.75 \\ \midrule %noedge_con 
            T+E(A) & 0.79 & {\bf 0.82} & 0.81 & {\bf 0.69} & 0.65 & {\bf 0.67} & {\bf 0.74} & {\bf 0.76} \\ %noedge_ave
            T+E(C) & 0.80 & {\bf 0.82} & 0.81 & {\bf 0.69} & {\bf 0.66} & {\bf 0.67} & {\bf 0.74} & {\bf 0.76} \\ %noedge_con
            I+E(A) & 0.76 & {\bf 0.82} & 0.79 & 0.66 & 0.58 & 0.61 & 0.70 & 0.73 \\ %noedge_ave
            I+E(C) & 0.76 & {\bf 0.82} & 0.79 & 0.66 & 0.59 & 0.62 & 0.70 & 0.73 \\ %noedge_con
            T+I+E(A) & 0.78 & 0.81 & 0.80 & 0.68 & 0.63 & 0.65 & 0.73 & 0.74 \\ %ave
            T+I+E(C) & 0.79 & {\bf 0.82} & 0.80 & 0.68 & 0.64 & 0.66 & 0.73 & 0.75 \\ %con
            T+I+E(A+C) & 0.78 & 0.81 & 0.80 & 0.67 & 0.63 & 0.65 & 0.72 & 0.74 \\ 
            \bottomrule
        \end{tabular}
        \label{tab/Ablation_Result}
    \end{center}
\end{table}
We compare the performance of the proposed models described in Section \ref{subsubsec:Compare_Originalmodel}  
with the model from previous research \cite{sosea2021Using}.  
The precision, recall, and F1 score for ``Similar'' and ``Complementary'' labels in DisRel,  
as well as the Macro-F1 score and accuracy, are calculated,  
and the results are summarized in Table \ref{tab/Ablation_Result}.  
The values for each model represent the average of 10 trials,  
where P indicates precision and R indicates recall.  
Additionally, T, I, and E denote the use of text, image, and edge vectors  
(ITRC-Line Graph node vectors), respectively.  
For the existing models, 
we compare our approach with two models: 
ViLBERT-REL-MT \cite{lu2019ViLBERTa} and models using Edge2vec \cite{wang2020Edge2vec}. 
The former model demonstrated the highest performance by incorporating a help task from a different dataset, 
as reported in \cite{sosea2021Using}. 
The result of ViLBERT-REL-MT is quoted from \cite{sosea2021Using}.
The latter model is evaluated in the same experiment as the proposed method.

First, 
comparing the proposed models with the existing model,  
we find that the method of concatenating text and edge vectors using T+E(C)
achieves the highest performance.  
% We call this T+E(C).  
Notably, 
compared to the existing best model, ViLBERT-REL-MT,  
T+E(C) improves precision by 6 points,  
recall by 4 points, and F1 score by 5 points for the ``Complementary'' label,  
along with a 2-point increase in Macro-F1.  
This significant improvement demonstrates a substantial enhancement in  
the accuracy of detecting complementary relationships.  
Moreover, while maintaining the performance for ``Similar'' relationships,  
the proposed model also contributes to the overall task performance.  
Additionally,
T+E(C) consistently outperform those using Edge2Vec \cite{wang2020Edge2vec}.
However, the overall accuracy remains comparable to existing methods,  
indicating the need for further improvement in comprehensive performance.  

Next, 
we compare the baseline and proposed models in this study.  
Among the baseline models, 
the method of concatenating image and text vectors  
(T+I(C)) exhibits higher performance than averaging. 
% identifying it as T+I(C).  
Both the best baseline 
and the proposed best models (T+E(C)) outperform the highest-performing  
existing model in terms of ``Complementary'' and Macro-F1 scores,  
suggesting that the use of CLIP encoders effectively contributes to  
the detection of ``Complementary'' labels.  
Furthermore, 
T+E(C) improves recall for the ``Complementary'' label  
by 3 points and Macro-F1 by 1 point compared to T+I(C).  
This result indicates that leveraging edge information contributes to performance improvement,  
highlighting that, in addition to the internal relationships of image-text pairs,  
considering external relationships through clustering and ITRC-Line Graph  
effectively enhances the accuracy of detecting complementary relationships.  

Finally, we analyze the impact of different fusion methods on performance  
through comparisons within the proposed models.  
The results show that concatenation methods exhibit slightly better performance  
than averaging methods, likely due to preserving the original embeddings.  
However, when fusing all three vectors (text, image, and edge),  
the performance tends to decline compared to fusing only text and edge vectors.  
This observation suggests that text features are more critical than image features  
in identifying complementary relationships.  

In summary, 
our method fusing text and edge vectors significantly outperforms  
existing methods in detecting ``Complementary'' relationships,  
while the overall performance remains comparable to previous approaches.

% \ref{subsubsec:Compare_Originalmodel}節で提案したモデル間および先行研究のモデル\cite{sosea2021Using}の性能を比較する．
% DisRelの``Similar''と``Complementary''に対する適合率，再現率，F値，
% またMacro-F1と正解率を算出した結果を表\ref{tab/Ablation_Result}に示す．
% 各モデルの値は10回の実験の平均値で，
% 表中のPは適合率，
% Rは再現率を表す．
% またTがテキスト，
% Iが画像，
% Eがエッジベクトル(ITRC-Line Graphのノードベクトル)の使用を指す．
% また先行研究のモデルとして，
% 最も性能が良いと示されているViLBERT\cite{lu2019ViLBERTa}に別のデータセットによるヘルプタスクを加えたモデルの結果を論文\cite{sosea2021Using}から引用する．

% まず既存モデルと提案モデルを比較した結果，
% テキストとエッジを方法2で連結する手法が最も高い性能を示した．
% これを提案ベストモデルと呼ぶ．
% 特に，
% 既存の最高性能モデルViLBERT-REL-MTと比較して，
% ``Complementary''の適合率で6ポイント，
% 再現率で4ポイント，
% F1値で5ポイント，
% Macro-F1で2ポイント向上し，
% 補足関係の判別精度の大幅な性能向上に成功している．
% また，
% ``Similar''の判別性能を維持しつつ，
% タスク全体の性能向上にも寄与している．
% しかし，
% 全体的な正答率は既存手法と同等であり，
% 更なる全体性能の改善は課題として残されている．

% 次に本研究のベースラインモデルと提案モデルを比較する．
% まずベースラインのモデルは，
% 画像とテキストを方法2の連結での融合の性能が平均での融合よりも高く，
% このモデルがベストベースラインモデルである．
% そしてベストベースラインおよび提案ベストモデルは，
% 共に``Complementary''やMacro-F1の各評価値で先行研究の最高性能を上回り，
% CLIPのエンコーダーが``Complementary''ラベルの判別に寄与していることが示唆された．
% 更に，
% 提案ベストモデルはベストベースラインモデルと比べて，
% ``Complementary''の再現率を3ポイント，
% Macro-F1を1ポイント向上させている．
% この結果は，
% エッジ情報の活用が性能向上に寄与していることを示し，
% 画像ーテキスト対の内部関係に加え，
% クラスタリングやITRC-Line Graphによる対の外部関係の考慮が，
% 補足関係の判別精度向上に有効であると考えられる．

% 最後に提案モデル内の比較により，
% 融合方法の違いが性能に与える影響を分析する．
% その結果，
% 平均手法よりも連結手法の方がやや高い性能を示しており，
% 各エンべディングの表現をそのまま活用できることが要因と考えられる．
% しかし，
% 3種類全てのベクトルを融合した場合，
% テキストベクトルとエッジベクトルのみを融合した場合に比べて性能が低下する傾向が確認された．
% このことから，
% テキストが画像より判別において重要な要素である可能性が示唆された．

% 以上から，
% テキストベクトルとエッジベクトルの融合により``Complementary''ラベルの判別精度が既存手法を大幅に上回ったものの，
% 全体的な性能は既存手法と同程度であり，
% 更なる改善の課題が残されている．

\section{Discussion}
\label{sec:Analysis}
First, the distribution of each embedding per label by CLIP of DisRel images and texts is analysed (Section \ref{subsec:Analysis_Dataset}).
Next, the performance of the representation learning of the ITRC-Line Graph is evaluated (Section \ref{subsec:Analysis_ITRC-LineGraph}).
Finally, 
as a case study, 
a qualitative analysis is performed by comparing the actual outputs (Section \ref{subsec:Analysis_qualitative}).

% まずDisRelの画像とテキストのCLIPによるラベル毎の各エンべディングの分布を分析する(\ref{subsec:Analysis_Dataset}節)．
% 次にITRC-Line Graphの表現学習の性能評価を行う(\ref{subsec:Analysis_ITRC-LineGraph}節)．
% 最後にケーススタディとして実際の出力の比較により定性的分析を行う(\ref{subsec:Analysis_qualitative}節)．

\subsection{Analysis of Dataset}
\label{subsec:Analysis_Dataset}
\begin{figure}[t]
    \centering
    \includegraphics[width=0.8\linewidth]{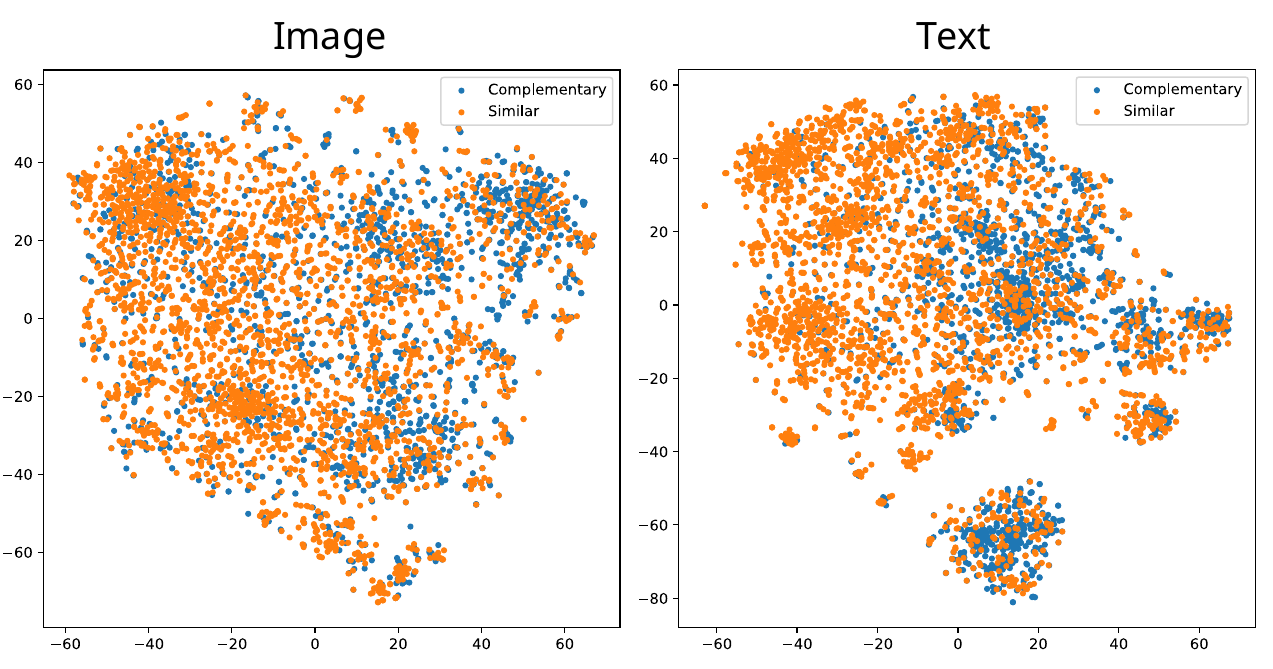}
    % \caption{DisRelの``Complementary''(青色)と``Similar''(橙色)の全4700個のデータに対する，
    % CLIPのエンコーダによる次元削減後のエンべディングの分布(左が画像で，右がテキスト)．}
    \caption{Distribution of image and text embeddings after dimensionality reduction by the CLIP encoder. 
    ``Complementary'' and ``Similar'' class in DisRel are illustrated in blue and orange, respectively.}
    \label{fig:fig_OriginalVectors}
    \vspace{-2em}
\end{figure}
To visualize and analyze the DisRel dataset,  
we embedded images and texts into 512-dimensional vectors using the CLIP encoder  
and then reduced the dimensionality to 2D using t-SNE \cite{maaten2008Visualizing}.  
We plotted 4,700 pairs labeled as either ``Complementary'' or ``Similar''  
by coloring them differently according to their labels.  

The result is shown in Figure \ref{fig:fig_OriginalVectors}.  
While no significant difference was observed in the distribution of images,  
the text vectors revealed a high-density region corresponding to  
the ``Complementary'' label.  
This observation suggests that text features may play a crucial role  
in relation classification.  

% DisRelデータセットを可視化・分析するため，
% CLIPのエンコーダで画像とテキストを512次元にエンべディングし，
% t-SNEで2次元に次元削減した．
% そして，
% ``Complementary''と``Similar''の2種類のラベルの計4700ペアをラベル別に異なる色でプロットした．

% 結果は図\ref{fig:fig_OriginalVectors}である．
% 画像の分布には大きな差異が見られなかったが，
% テキストには一部``Complementary''について高密度な領域が確認される．
% これにより，
% 関連性の分類にはテキストが重要である可能性を示唆している．

\subsection{Analysis of ITRC-Line Graph Representation Learning}
\label{subsec:Analysis_ITRC-LineGraph}
\begin{table}[t]
    \centering
    \caption{Statistics of ITRC-Line Graph}
    \begin{tabular}{ccccccc} \toprule
        \multicolumn{2}{c}{\textbf{The number of edge}} & \multicolumn{5}{c}{\textbf{The number of node($L$)}} \\ \midrule
        Total & \multirow{2}{*}{\begin{tabular}{c} after reduction  \\ (Average) \end{tabular}} &
        Total & \multirow{2}{*}{\begin{tabular}{c} Training \\ (Average) \end{tabular}} & \multicolumn{3}{c}{\begin{tabular}{c} Test(Average) \end{tabular}} \\ \cmidrule{5-7}
        & & & & Total & Sim & Com \\ \midrule
        48702 & 35925 & 1928 & 1409 & 519 & 336 & 183 \\ \bottomrule
    \end{tabular}
    \label{tab/ITRC-LineGraph_information}
\end{table}
In Section~\ref{subsubsec:ITRC-Line-Graph}, 
we evaluate the performance of representation learning for the ITRC-Line Graph $G^*$.  
This evaluation analyzes the subtasks related to edge vector learning presented in Section~\ref{subsec:result}, 
using the average results of 10 trials.  
It should be noted that the original pair data $P$ and the labeling method differ.  

In this experiment, we utilize the ITRC-Line Graph $G^*$ after edge reduction.  
Following the procedure described in Section~\ref{secsubsub:Model_Setup}, we split the graph into training and test nodes.  
The proportion of training data averaged 0.73, and the detailed information is shown in Table~\ref{tab/ITRC-LineGraph_information}.  

% \ref{subsubsec:ITRC-Line-Graph}節でのITRC-Line Graph $G^*$ の表現学習の性能を評価する．
% 本評価では，
% \ref{subsec:result}節に示すエッジベクトルの学習に関するサブタスクを分析し，
% 10 回の平均結果を用いる．
% また，
% 元のペアデータ $P$ とラベル付けの方法が異なる点に留意する．
% 本実験では，
% エッジ削減後の ITRC-Line Graph $G^*$ を用いており，
% \ref{secsubsub:Model_Setup}節に従い，
% 訓練ノードとテストノードに分割した．
% この時訓練データ割合は平均で0.73であり，
% その詳細な構成については表\ref{tab/ITRC-LineGraph_information} に示す．

\begin{table}[t]
  \centering
  \caption{Classification Results for ITRC-Line Graph (rounded to the third decimal place, average of 10 times).}
  \begin{tabular}{cccccccc} \toprule
      \multicolumn{3}{c}{Similar} & \multicolumn{3}{c}{Complementary} & Macro F1 & Accuracy \\ \cmidrule{1-6}
      P & R & F1 & P & R & F1 & & (Acc)\\ \midrule
      0.72 & 0.88 & 0.79 & 0.63 & 0.39 & 0.48 & 0.64 & 0.70 \\
      \bottomrule
  \end{tabular}
  \label{tab/ITRC-LineGraph_Result}
\end{table}
The experimental results are shown in Table~\ref{tab/ITRC-LineGraph_Result}.  
The accuracy is $70\%$, and the Macro-F1 score is 0.64. The F1 score is not high due to the low recall for the ``Complementary'' label,  
while the recall and F1 score for ``Similar'' are higher.  
On the other hand, 
by combining the edge vectors with the original pairs, 
we observed an improvement in the performance for the ``Complementary'' label.  
Out of the dataset $P$ (4700 pairs), 
about 82\% (on average, 3853 pairs) of the labels on the ITRC-Line Graph matched.  
Furthermore, following the labeling rules in Section~\ref{secsubsub:Model_Setup}, some test data were learned as training nodes, even though they were not included in the majority vote,  
indicating that some of the test data were effectively learned.

From these results,
it can be suggested that aggregating image and text embeddings at the nodes of the ITRC-Line Graph,
combined with majority voting for labeling, 
is effective.
This approach considers the internal and external interactions between image-text pairs,
particularly contributing to the discrimination of the “Complementary” label.
% From these results, it can be suggested that by aggregating image and text embeddings at the nodes of the ITRC-Line Graph, and applying majority voting for labeling,  
% the internal and external interactions between image-text pairs are considered, which particularly contributes to the discrimination of the ``Complementary'' label.  
Additionally, it is likely that further fusion of the text vector with the edge vectors formed from the image and text is also important.

% 実験結果を表\ref{tab/ITRC-LineGraph_Result}に示す．
% 正答率は $70\%$，Macro-F1 値は 0.64 であり，
% ``Complementary''のラベルについての再現率が低いためF1値は高くなく，
% ``Similar''に対して再現率が高くF1値が高いという結果となっている．
% 一方で，
% 元のペアに対しエッジベクトルを結合することで``Complementary''の性能が改善したということが示されている．
% ここでデータセット $P$ (4700 ペア) のうち，
% ITRC-Line Graph上でのラベル付けと約82\%(平均3853ペア)が一致するラベルを持っていた．
% また，
% ITRC-Line Graphの構成時に，
% \ref{secsubsub:Model_Setup}節のラベリングルールに則ると，
% 一部のテストデータは多数決に含まれていなくとも訓練ノードとして学習されており，
% テストデータの一部が効果的に学習されている可能性がある．

% 以上の結果から，
% ITRC-Line Graph のノードにおいて，
% 画像・テキストのエンべディングを集約し，
% 多数決によるラベル付けを適用することで，
% 画像ーテキスト対の内部および外部の相互作用を考慮し，
% 特に``Complementary'' の判別に寄与している可能性が示唆されている．
% 加えて，
% 画像とテキストから構成されたエッジベクトルに対して，
% テキストベクトルをもう一度融合することも重要であると考えられる．

\subsection{Case Study}
\label{subsec:Analysis_qualitative}
We analyzed the output of 10 test data samples for T+E(C) (T+E (Concatenation)) and the baseline model (T+I (Concatenation)), comparing five examples. 
Additionally, 
we compared the results with ChatGPT's output. 
The relevance definitions from Section~\ref{sec:intro} were provided as prompts to ChatGPT. 
The GPT-4o model (as of January 2025) was used for this experiment \cite{chatgpt2025}. 
The output results are shown in Table~\ref{fig_casestudy_compare}.
\begin{itemize}[left=0em, label={}]
    \item \textbf{Pair I:} 
    The text and image describe the impact of Hurricane Irma across Florida. 
    All models correctly identified this as a complementary relationship.
    \item \textbf{Pair II:} 
    The text describes the weakened Hurricane Irma, 
    and the image shows its intensity. 
    The baseline model misclassified this as a similar relationship, 
    while the proposed model and ChatGPT correctly identified it as complementary.
    \item \textbf{Pair III:} 
    The text discusses storm surges one week after Irma, 
    and the image shows rising water levels. 
    This should be a similar relationship, 
    but the baseline model and ChatGPT misclassified it. 
    The proposed model performed better.
    \item \textbf{Pair IV:} 
    The text describes a flood warning in South Carolina, 
    and the image shows the flood. 
    The proposed model misclassified this as a similar relationship, 
    while ChatGPT correctly identified it as complementary by focusing on the warning aspect.
    \item \textbf{Pair V:} 
    Although expected to be complementary from the perspective of city officials, 
    all models classified it as similar. 
    This highlights a potential issue with proper noun entities, 
    as noted in previous research \cite{sosea2021Using}, 
    suggesting the need for relabeling in the dataset.
\end{itemize}

% \paragraph{Pair I:} The text and image describe the impact of Hurricane Irma across Florida. All models correctly identified this as a complementary relationship.

% \paragraph{Pair II:} The text describes the weakened Hurricane Irma, and the image shows its intensity. The baseline model misclassified this as a similar relationship, while the proposed model and ChatGPT correctly identified it as complementary.

% \paragraph{Pair III:} The text discusses storm surges one week after Irma, and the image shows rising water levels. This should be a similar relationship, but the baseline model and ChatGPT misclassified it. The proposed model performed better.

% \paragraph{Pair IV:} The text describes a flood warning in South Carolina, and the image shows the flood. The proposed model misclassified this as a similar relationship, while ChatGPT correctly identified it as complementary by focusing on the warning aspect.

% \paragraph{Pair V:} Although expected to be complementary from the perspective of city officials, all models classified it as similar. This highlights a potential issue with proper noun entities, as noted in previous research \cite{sosea2021Using}, suggesting the need for relabeling in the dataset.

These analyses reveal differences in the focus between the image and text, 
as well as challenges with labels, 
emphasizing important considerations when embedding multimodal data into a common space.

\begin{table}[t]
  \centering
  \caption{Results for Each of the Relevance Prediction Labels.}
  \label{fig_casestudy_compare}
  \begin{tabular}{c|ccccc} 
      \toprule
      & I & I\hspace{-1.2pt}I & I\hspace{-1.2pt}I\hspace{-1.2pt}I & I\hspace{-1.2pt}V & V \\ 
      \midrule
      Image & 
      \includegraphics[width=0.16\linewidth]{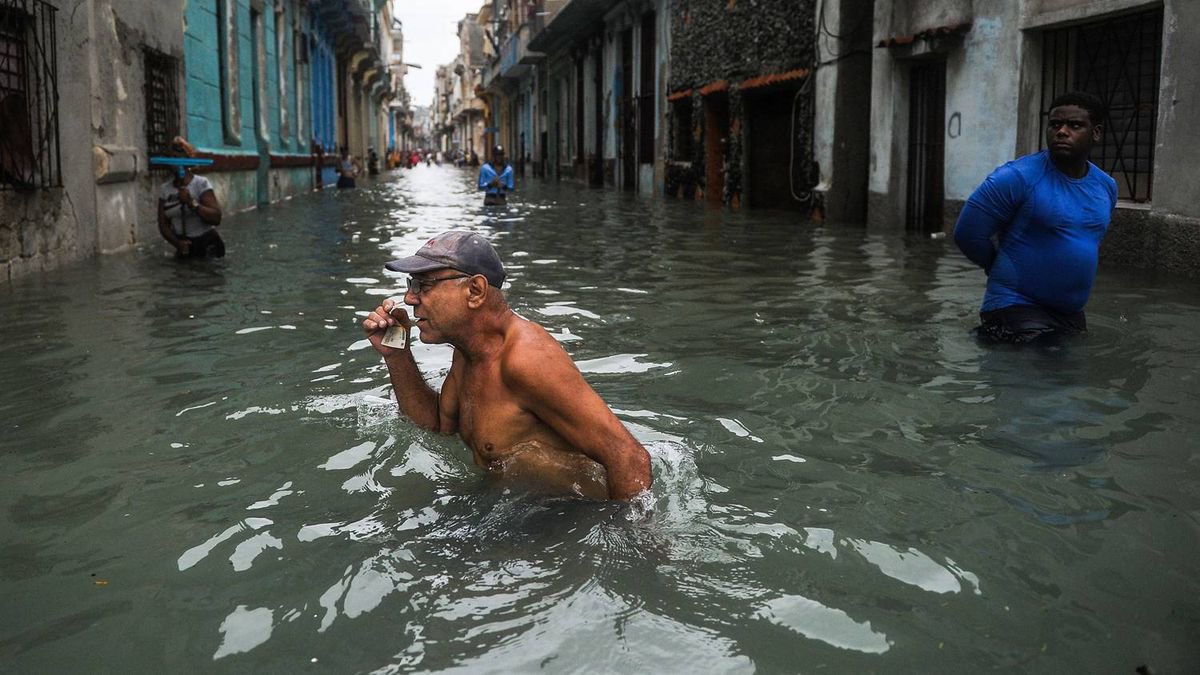} &
      \includegraphics[width=0.16\linewidth]{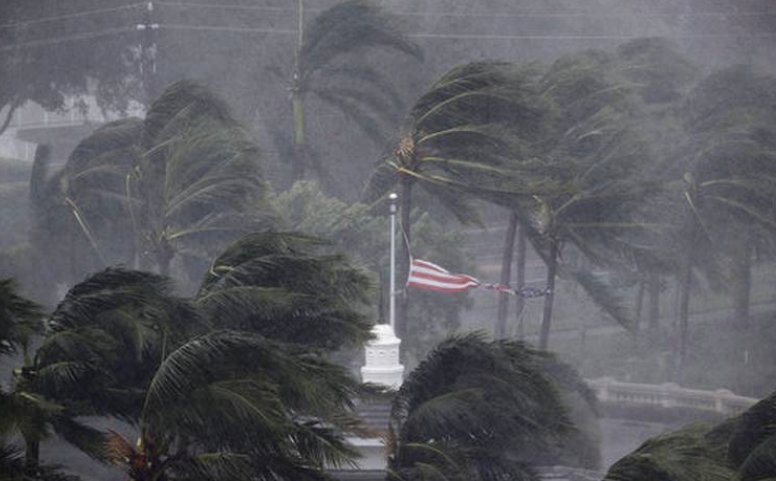} &
      \includegraphics[width=0.16\linewidth]{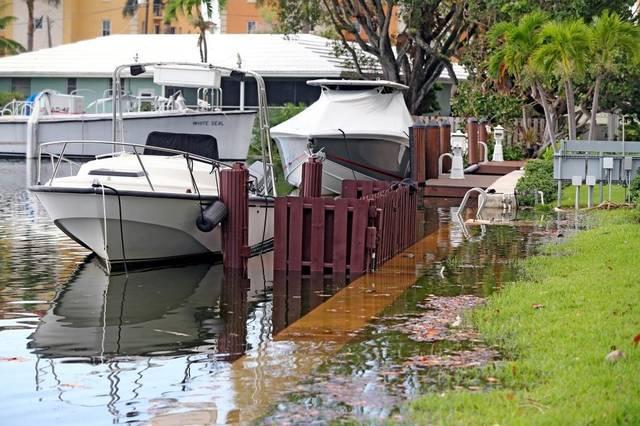} &
      \includegraphics[width=0.16\linewidth]{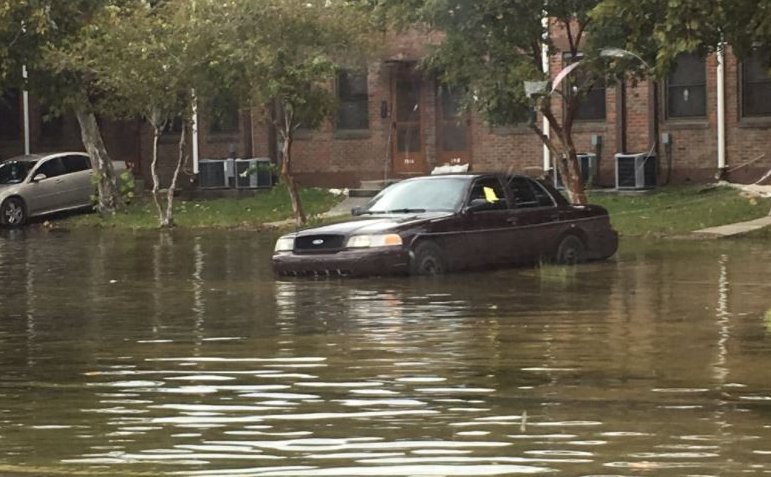} &
      \includegraphics[width=0.16\linewidth]{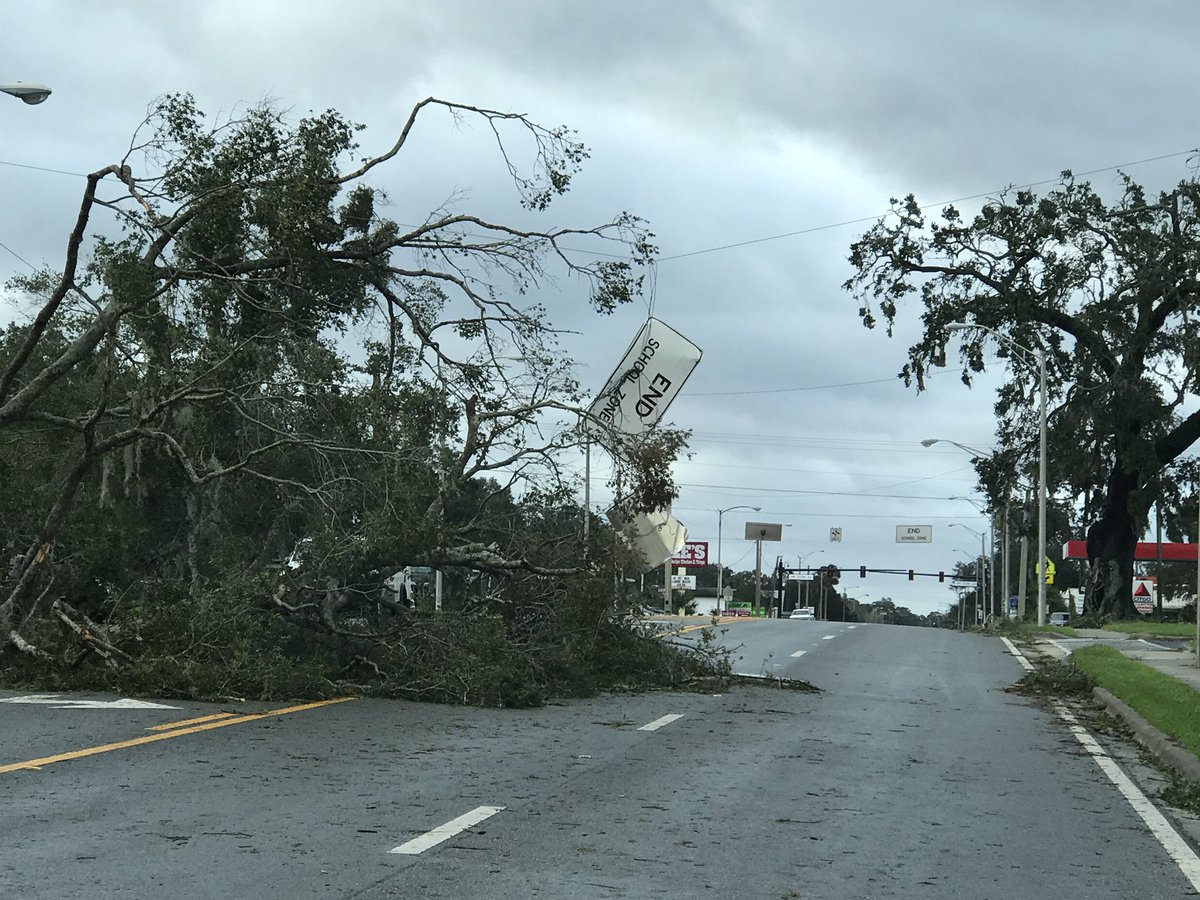} \\
      Text & 
      \parbox{0.15\linewidth}{\centering \scriptsize "Hurricane Irma downgraded to Category 1, but still wreaking havoc across Florida <URL> \#Irma <URL>"} &
      \parbox{0.15\linewidth}{\centering \scriptsize Weakened \#Irma lashes much of Fl <URL>} &
      \parbox{0.15\linewidth}{\centering \scriptsize King tides rise a week after Irma, highlighting flood risks <URL> @entornoi <URL>} &
      \parbox{0.15\linewidth}{\centering \scriptsize \#HurricaneIrma: Flooding in South Carolina as officials warn residents remain alert <URL> <URL>} &
      \parbox{0.15\linewidth}{\centering \scriptsize "A tree is down on 17-92 near W 18th St. in Sanford, blocking most of the road. City crew on scene \#Irma \#WFTV <URL>"} \\ 
      \midrule
      \bf{Correct} & \bf{Com} & \bf{Com} & \bf{Sim} & \bf{Com} & \bf{Com} \\ 
      T+I(C) & \bf{Com} & Sim & Com & \bf{Com} & Sim \\ 
      ChatGPT & \bf{Com} & \bf{Com} & Com & \bf{Com} & Sim \\
      T+E(C) & \bf{Com} & \bf{Com} & \bf{Sim} & Sim & Sim \\ 
      \bottomrule
  \end{tabular}
\end{table}

\section{Conclusion}
\label{sub:Conc}
In this paper, 
we highlighted the importance of extracting complementary relationships in the classification of image-text pairs in SNS posts and proposed a graph-based method for automatic classification. 
Specifically, 
we introduced a method that integrates GCN-based embeddings obtained from the ITRC-Line Graph with the original image-text pair embeddings 
and classifies them using an MLP classifier. 
Experimental results showed that the F1 score for the complementary relationship improved to $0.67$, 
significantly outperforming existing methods.
% The overall classification accuracy remained at $76\%$. 
A case study analysis confirmed the effectiveness of the proposed method in distinguishing between complementary and similar relationships. 
Future challenges include optimizing graph learning and fusion techniques to further improve classification performance.
Additionally,
we plan to evaluate the generalizability and applicability of our method using different datasets.

\renewcommand{\baselinestretch}{0.973}
\bibliographystyle{main}
\bibliography{main}
%
% \begin{thebibliography}{8}
% \bibitem{ref_article1}
% Author, F.: Article title. Journal \textbf{2}(5), 99--110 (2016)

% \bibitem{ref_lncs1}
% Author, F., Author, S.: Title of a proceedings paper. In: Editor,
% F., Editor, S. (eds.) CONFERENCE 2016, LNCS, vol. 9999, pp. 1--13.
% Springer, Heidelberg (2016). \doi{10.10007/1234567890}

% \bibitem{ref_book1}
% Author, F., Author, S., Author, T.: Book title. 2nd edn. Publisher,
% Location (1999)

% \bibitem{ref_proc1}
% Author, A.-B.: Contribution title. In: 9th International Proceedings
% on Proceedings, pp. 1--2. Publisher, Location (2010)

% \bibitem{ref_url1}
% LNCS Homepage, \url{http://www.springer.com/lncs}, last accessed 2023/10/25
% \end{thebibliography}
\end{document}